\documentclass{article}
\usepackage{bbm}

 \usepackage[preprint]{neurips_2026}

\usepackage[utf8]{inputenc} 
\usepackage[T1]{fontenc}    
\usepackage{hyperref}       
\usepackage{url}            
\usepackage{booktabs}       
\usepackage{amsfonts}       
\usepackage{nicefrac}       
\usepackage{microtype}      
\usepackage{xcolor}         
\usepackage{graphicx}
\usepackage{colortbl}
\definecolor{tblHeader}{HTML}{E8EEF7}
\definecolor{tblGroupA}{HTML}{F7FAFF}
\definecolor{tblGroupB}{HTML}{FFF8E8}
\definecolor{tblGroupC}{HTML}{EAF7EA}
\definecolor{darkgreen}{HTML}{0B6E2E}
\definecolor{darkred}{HTML}{9B1C1C}
\usepackage[most]{tcolorbox}

\definecolor{abstractblue}{HTML}{F1F7FF}

\title{Benchmarking the Benchmarks: A Validity Audit of Tool-Calling Evaluation}

\author{%
  Jay Vaghasiya$^{1}$, Vishvesh Bhat$^{1}$, Muhammad Ahmed Mohsin$^{2}$, Asad Aali$^{2}$\\[0.5em]
  $^{1}$CoreThink AI \\
  $^{2}$Stanford University
}

\begin{document}

\maketitle

\begin{center}
\begin{minipage}{0.96\linewidth}
\begin{tcolorbox}[
    colback=abstractblue,
    colframe=abstractblue,
    boxrule=0pt,
    arc=2pt,
    left=7pt,
    right=7pt,
    top=6pt,
    bottom=6pt,
    enhanced,
    breakable
]
\textbf{Abstract.}
Tool-calling benchmarks are increasingly used to rank language-model agents, yet their scores are often treated as ground truth without validating the evaluators themselves. We present a systematic validity and reproducibility audit of four major tool-calling benchmark families: BFCL v4, $\tau^2$-Bench, LiveMCPBench, and MCP-Atlas. Across 496 expert-reviewed benchmark tasks, we find 92 evaluator-human disagreements, corresponding to an 18.5\% misalignment rate. The failures are not isolated annotation mistakes: deterministic benchmarks exhibit brittle state matching, trajectory lock-in, incorrect ground truths, substring-based communication failures, and reward-basis misalignment, while LLM-judge benchmarks exhibit rubric drift, hallucinated completion, answer-only scoring, and substantial run-to-run variance. In LiveMCPBench, 23 repeated evaluations of the same setup produce scores ranging from 57.9\% to 76.8\%, a spread of 18.9 percentage points, large enough to change leaderboard conclusions. These results show that current tool-calling scores can reflect evaluator artifacts rather than agent capability. We introduce a unified taxonomy of tool-calling evaluation failures, release trace-level audit artifacts and corrected evaluation components, and argue for decomposed metrics that separately measure tool invocation, task completion, and outcome verification. Our findings suggest that progress in tool-using agents requires benchmarks whose evaluators are themselves reproducible, auditable, and aligned with human judgments of task success. We further introduce \textbf{Tool-Veritas}, a configurable benchmark that combines deterministic state verification with optional qualitative judging, and \textbf{Harness Lab}, an open-source system for benchmark execution, trace inspection, repeated-run comparison, and evaluator debugging.
\end{tcolorbox}
\end{minipage}
\end{center}

\section{Introduction}

Tool-calling has emerged as a core capability of modern language-model agents, enabling interaction with external APIs, databases, operating systems, web services, and enterprise tools~\citep{patil2023gorilla,li2023apibank,qin2024toolllm,guo2024stabletoolbench}. Consequently, benchmark-driven evaluation has become the primary mechanism for measuring progress, comparing models, and guiding deployment decisions. Benchmarks such as BFCL, $\tau$-bench, $\tau^2$-Bench, LiveMCPBench, and MCP-Atlas are increasingly used as standard references for function calling, stateful interaction, and large-scale tool-use evaluation~\citep{berkeleyfunctioncallleaderboard,yao2024taubench,barres2025tau2bench,mo2025livemcpbench,bandi2026mcpatlas}. Implicit in their adoption is the assumption that benchmark scores provide a reliable measurement of tool-calling capability.

However, benchmark validity requires more than task coverage and scale. An evaluation must correctly distinguish successful from unsuccessful executions, faithfully capture user intent, and produce reproducible scores under repeated evaluation. Prior work has already shown that tool-use and LLM evaluation can be sensitive to API instability, automatic evaluator design, judge bias, and preference-aggregation artifacts~\citep{guo2024stabletoolbench,zheng2023judging,dubois2024lengthcontrolled,chiang2024chatbotarena}. We find that many existing tool-calling benchmarks fail to satisfy these requirements in more fundamental ways. Through a systematic audit of four widely used benchmarks, we identify pervasive evaluation artifacts that are often unrelated to actual tool-use ability. Deterministic benchmarks suffer from brittle state comparisons, exact-match constraints, incorrect ground-truth annotations, trajectory lock-in, and reward formulations that fail to capture task completion. LLM-judge benchmarks exhibit a different failure mode: stochastic rubric generation, judge variance, hallucinated evaluations, and implementation-paper mismatches that introduce substantial score instability.

These issues have practical consequences. We identify numerous false negatives where agents correctly complete tasks but are marked as failures due to evaluator artifacts, as well as false positives where benchmarks assign passing scores despite incomplete or incorrect task execution. We further show that leaderboard results can become highly unstable under repeated evaluation; for example, rerunning LiveMCPBench with its default evaluation pipeline produces score swings approaching twenty percentage points without changing the evaluated model. Collectively, these findings suggest that current benchmark scores often reflect properties of the evaluator rather than properties of the agent being evaluated.

To address these limitations, this work makes five contributions. \textbf{First}, we conduct a trace-level audit of 496 executions across BFCL v4, $\tau^2$-Bench, LiveMCPBench, and MCP-Atlas, identifying 92 evaluator--human disagreements. \textbf{Second}, we develop a unified taxonomy of deterministic and LLM-judge evaluation failures, including brittle state matching, trajectory lock-in, annotation errors, reward-basis mismatch, rubric drift, hallucinated completion, and judge variance. \textbf{Third}, we propose a decomposed evaluation framework that separately measures tool invocation, task completion, and outcome verification. \textbf{Fourth}, we introduce \textbf{Tool-Veritas}, a configurable tool-calling benchmark that evaluates factual task completion through deterministic state gates and uses an optional restricted LLM judge only for qualitative criteria. \textbf{Fifth}, we develop \textbf{Harness Lab}, a benchmark execution and debugging system that preserves raw artifacts, exposes case- and turn-level diagnostics, compares repeated runs, supports selective retry, and records human adjudications.

\section{Related Work}

\paragraph{Tool-use and function-calling benchmarks.}
Early tool-use benchmarks evaluate whether language models can select APIs, fill arguments, and produce valid tool calls, including ToolBench~\citep{xu2023toolbench}, ToolAlpaca~\citep{tang2023toolalpaca}, API-Bank~\citep{li2023apibank}, Gorilla~\citep{patil2023gorilla}, ToolLLM~\citep{qin2023toolllm}, and BFCL~\citep{patil2024bfcl}. These benchmarks substantially advanced standardized function-calling evaluation, but they often reduce success to syntactic call matching, AST equivalence, or narrow execution checks. As a result, they under-specify whether the agent actually satisfies user intent in realistic multi-turn environments.

\paragraph{Interactive and stateful agent evaluation.}
Recent benchmarks move beyond isolated tool calls toward stateful interaction, simulated users, and persistent environments, including $\tau$-bench~\citep{yao2024taubench}, $\tau^2$-bench~\citep{barres2025tau2bench}, ToolSandbox~\citep{lu2024toolsandbox}, WebArena~\citep{zhou2023webarena}, WorkArena~\citep{drouin2024workarena}, and OSWorld~\citep{xie2024osworld}. These settings better capture long-horizon tool use, policy following, and environment updates, but their scoring is typically tied to final database hashes, scripted trajectories, or brittle natural-language checks. This makes them vulnerable to false negatives when agents reach valid alternative outcomes and false positives when unchanged state is incorrectly treated as success. MAVEN introduces a verification-centered reasoning scaffold and MAVEN-Bench, an adversarial multi-step math and physics benchmark that exposes substantial cross-benchmark degradation in frontier models and highlights the need for process-aware evaluation of tool-using agents~\cite{maven}.

\paragraph{MCP and LLM-as-a-judge evaluations.}
MCP-based benchmarks such as LiveMCPBench~\citep{mo2025livemcpbench}, MCP-Bench~\citep{wang2025mcpbench}, MCP-Universe~\citep{luo2025mcpuniverse}, MCPMark~\citep{wu2025mcpmark}, and MCP-Atlas~\citep{bandi2026mcpatlas} evaluate agents in larger tool ecosystems with realistic multi-server workflows. To scale evaluation, many rely on LLM-as-a-judge protocols inspired by MT-Bench~\citep{zheng2023judging}, AlpacaEval~\citep{li2023alpacaeval}, and Chatbot Arena~\citep{chiang2024chatbotarena}. However, these evaluators often judge final answers rather than verified tool use, introduce rubric and judge variance, and can hallucinate task completion, leaving benchmark scores sensitive to evaluator artifacts rather than agent capability.

\paragraph{Benchmark execution and debugging infrastructure.}
Existing benchmark releases generally provide benchmark-specific execution scripts and aggregate scoring, but offer limited support for cross-benchmark trace inspection, repeated-run comparison, selective case retry, and persistent human adjudication. Harness Lab addresses this infrastructure gap by providing a common execution and debugging layer across heterogeneous benchmark implementations while preserving each benchmark's original evaluator and raw artifacts.

\section{Methodology}
\label{sec:methodology}

\subsection{Preliminaries}

We model a tool-calling benchmark as a tuple $\mathcal{B}=(\mathcal{T},\mathcal{E},\mathcal{G})$, where $\mathcal{T}$ denotes the benchmark tasks, $\mathcal{E}$ the evaluation harness, and $\mathcal{G}$ the benchmark-specific success criterion. Each task $t\in\mathcal{T}$ consists of a user query $q$, an initial environment state $s_0$, a set of available tools $\mathcal{A}$, and an expected outcome $o^{*}$. Given an agent $\pi$, task execution produces a trajectory $\tau=\{(s_0,a_1), (s_1,a_2), \ldots, (s_T,a_T)\}$ consisting of tool invocations, environment transitions, and natural-language responses. The benchmark evaluator assigns a binary outcome $y=\mathcal{E}(\tau)$.

The central assumption underlying benchmark evaluation is that the benchmark verdict matches the true task outcome. Let $\mathcal{H}(\tau)$ denote the expert trace-level assessment of task success.An evaluator agrees with the expert judgment on trajectory $\tau$ when

\begin{equation}
\mathcal{E}(\tau)=\mathcal{H}(\tau).
\end{equation}
Our objective is to systematically identify and characterize instances where this equality fails.

\subsection{Benchmark Audit Framework}

We audit four representative tool-calling benchmarks spanning both deterministic and LLM-judge-based evaluation paradigms: BFCL, $\tau^2$-Bench, LiveMCPBench, and MCP-Atlas. For each benchmark, we collect complete execution traces, tool invocation logs, evaluator outputs, environment states, and final benchmark verdicts. Each benchmark instance is represented as $x_i=(\tau_i,y_i,m_i)$, where $\tau_i$ denotes the execution trace, $y_i$ the benchmark-assigned label, and $m_i$ benchmark-specific metadata including evaluator logs, expected states, and judge outputs.

Our audit procedure consists of three stages: (i) reproducing benchmark evaluations, (ii) manually adjudicating benchmark outcomes, and (iii) performing root-cause analysis on benchmark disagreements. This enables direct comparison between benchmark-assigned labels and human-verified task outcomes.

\subsection{Human Adjudication}

For each flagged instance, annotators inspect the complete interaction trajectory, tool execution sequence, state transitions, and final user-visible outcome. A human label $h_i \in \{\texttt{PASS},\texttt{FAIL}\}$ is assigned based on whether the user objective was successfully achieved.

We define a benchmark disagreement whenever the benchmark label differs from the human assessment. More formally,

\begin{equation}
\delta_i=\mathbbm{1}[y_i \neq h_i].
\end{equation}

We further partition disagreements into false negatives ($y_i=0,h_i=1$) and false positives ($y_i=1,h_i=0$), enabling systematic quantification of evaluator failures across benchmarks.

\subsection{Failure Taxonomy}

Using manually verified disagreements, we construct a unified taxonomy of tool-calling evaluation failures. For deterministic benchmarks, we identify failures arising from exact-match constraints, state over-specification, trajectory lock-in, annotation errors, and reward misalignment. For LLM-judge-based benchmarks, we identify rubric drift, judge variance, hallucinated completion, and implementation-specification mismatches.

Each disagreement is assigned to one or more failure categories through trace-level analysis and evaluator inspection. This taxonomy provides a benchmark-agnostic framework for characterizing evaluation failures independently of any specific benchmark implementation.

\subsection{Reproducibility Analysis}

For benchmarks employing stochastic judges, a single benchmark score is insufficient to characterize evaluation reliability. We therefore execute repeated benchmark runs under identical settings and measure score variability. Given a set of benchmark scores $S=\{s_1,\ldots,s_K\}$ obtained from $K$ independent evaluations, we report the mean score $\mu$, standard deviation $\sigma$, and score spread $\Delta$.

The score spread,

\begin{equation}
\Delta=\max(S)-\min(S),
\end{equation}

captures the maximum leaderboard variation attributable solely to evaluation stochasticity. Large values of $\Delta$ indicate poor reproducibility and unstable benchmark rankings.

\subsection{Corrected Evaluation Framework}

Our audit reveals that existing benchmarks frequently collapse multiple aspects of agent behavior into a single binary score. We therefore decompose evaluation into three independent components: tool invocation correctness, task completion correctness, and outcome verification. Rather than assigning a single pass/fail label, evaluation is represented as

\begin{equation}
\mathbf{c}=
\left(
C_{\text{tool}},
C_{\text{task}},
C_{\text{outcome}}
\right),
\end{equation}

where each component evaluates a distinct aspect of agent behavior. This decomposition localizes evaluation failures, improves interpretability, and reduces sensitivity to benchmark-specific artifacts.

\subsection{Benchmark Corrections}

Using the identified failure cases, we construct corrected benchmark annotations, repaired evaluation logic, and revised evaluation harnesses. The corrected benchmark suite removes erroneous ground-truth labels, reduces brittle state comparisons, aligns evaluation procedures with benchmark specifications, and incorporates reproducibility-aware reporting. These corrected artifacts form the basis of the benchmark re-evaluation presented in the subsequent sections.

\subsection{Tool-Veritas: Deterministic-First Tool-Calling Evaluation}
\label{sec:custom_bench}

Our audit reveals complementary limitations in existing evaluation paradigms. Deterministic evaluators are reproducible but can reject semantically valid executions because of rigid syntax, trajectory, or state matching. LLM-based evaluators accommodate legitimate variation but can introduce rubric drift, unsupported success judgments, and score instability. Tool-Veritas addresses this trade-off through a deterministic-first protocol with a restricted LLM fallback.

A Tool-Veritas task is defined as
\[
t =
\left(
q,
s_0,
\mathcal{A},
\mathcal{G},
\mathcal{J}
\right),
\]
where $q$ is the user request, $s_0$ is the initial sandbox state, $\mathcal{A}$ is the available tool set, $\mathcal{G}$ is a collection of deterministic state predicates, and $\mathcal{J}$ is an optional qualitative rubric.

After each interaction turn, deterministic gates inspect observable properties of the sandbox state. These gates verify factual outcomes such as whether a file was created, a database record was updated, a calendar event was added, or a required tool-mediated action was completed. For environment state $s_t$ after turn $t$, deterministic completion is defined as
\[
C_{\mathrm{state}}(t)
=
\prod_{j=1}^{m_t}
\mathbbm{1}
\left[
g_{t,j}(s_t)=1
\right],
\]
where $g_{t,j}$ is the $j$th required predicate and $m_t$ is the number of predicates evaluated at turn $t$. A task cannot pass factual-completion evaluation when a required gate fails.

The LLM judge is invoked only for criteria that cannot be represented reliably as state predicates, including communication quality, policy adherence, and the adequacy of user-facing confirmations. It cannot override a failed deterministic requirement. This separation prevents fluent responses from receiving credit when the required environment change did not occur, while retaining flexibility for genuinely qualitative criteria.

Tool-Veritas also provides a bounded repair window. When a tool invocation fails, the agent may inspect the returned error and issue a corrected action before the turn is finalized. First-attempt completion and completion after repair are recorded separately, preserving execution errors while measuring recovery.

Tools, domains, state predicates, and Model Context Protocol (MCP) endpoints are configuration-defined. Each run produces a per-turn JSONL trace containing the model action, tool response, state transition, gate result, repair status, and optional judge output. The current benchmark spans sixteen domains, including filesystem operations, databases, version control, banking, travel, calendars, smart-home control, and healthcare workflows.

The empirical analysis in Section~\ref{sec:rq4} shows that this evaluator achieves 95.5\% aggregate agreement with expert trace-level judgments across the evaluated model--task pairs. We therefore study deterministic verification with restricted LLM fallback as a general scoring design for agentic benchmarks, while leaving the detailed agreement analysis and cross-benchmark comparison to RQ4.

\begin{table*}[t]
\centering
\small
\caption{Evaluator characteristics of Tool-Veritas and the benchmark versions audited in this work.}
\label{tab:custom_bench_comparison}
\setlength{\tabcolsep}{4pt}
\renewcommand{\arraystretch}{1.12}

\begin{tabular}{
    p{0.15\textwidth}
    p{0.17\textwidth}
    p{0.17\textwidth}
    p{0.18\textwidth}
    p{0.24\textwidth}
}
\toprule
\textbf{Dimension}
&
\textbf{BFCL}
&
\textbf{$\tau/\tau^2$}
&
\textbf{MCP benchmarks}
&
\textbf{Tool-Veritas}
\\
\midrule

Primary evaluator
&
AST and simulator-state matching
&
Database checks and language assertions
&
LLM trajectory or claim judging
&
Deterministic state gates with restricted LLM fallback
\\

State verification
&
Simulator-state conformity
&
Domain-specific database checks
&
Primarily trajectory or claim coverage
&
Sandbox predicates over observable task outcomes
\\

Qualitative scoring
&
Not primary
&
Natural-language assertions
&
LLM-based judging
&
Applied only after required deterministic gates pass
\\

Recovery
&
No separate repair metric in the audited setting
&
Environment dependent
&
Limited explicit repair accounting
&
Bounded repair window with separate repair outcomes
\\

Extensibility
&
Fixed tasks and schemas
&
Fixed domains
&
Fixed server collections
&
Configuration-defined tools, domains, predicates, and MCP endpoints
\\

Trace output
&
Benchmark-specific logs
&
Interaction and reward logs
&
Trajectory and judge logs
&
Per-turn action, state, gate, repair, and judge records
\\

\bottomrule
\end{tabular}
\end{table*}

\subsection{Harness Lab}
\label{sec}

Harness Lab is the execution and debugging infrastructure used for the benchmark audit. It runs benchmark suites against OpenAI-compatible model endpoints, preserves raw benchmark artifacts, parses case-level results, and provides benchmark-specific diagnostic views.

The current implementation supports seven benchmark families across 21 suite variants: BFCL v3/v4, $\tau^2$-Bench, Toolathlon, MCP-Atlas, SWE-Bench Lite, LiveCodeBench v6, and LiveMCPBench. Each benchmark remains associated with its original runner and evaluator.

For every run, Harness Lab stores the raw harness output, official score files, inference logs, case summaries, turn-level diagnostics, and execution metadata. Raw artifacts are retained in object storage, while parsed records are materialized into a relational database for case-level queries.

Harness Lab provides benchmark-specific diagnostics. For BFCL, it identifies the first failed turn, error category, and state-field mismatch. For $\tau^2$-Bench, it exposes database checks, action assertions, communication assertions, and reward components. For MCP-Atlas and LiveMCPBench, it displays expected and observed tool trajectories together with judge outputs.

The system also supports paired run comparison. Cases are aligned by identifier and classified as passed in both runs, failed in both runs, improved in one run, or missing from one run. For multi-turn tasks, the comparison identifies the first turn at which two runs diverge.

Selective retry allows transiently failed cases to be regenerated without rerunning the complete benchmark. Retried cases are merged with unaffected results, after which the official evaluator is executed again. Human verdicts and unreliable-case annotations are stored separately from official benchmark labels, preserving the distinction between automated scores and expert adjudication.

\begin{table}[t]
\centering
\small
\caption{Harness Lab functionality used in the audit.}
\label{tab:harness_lab_functionality}
\setlength{\tabcolsep}{5pt}
\renewcommand{\arraystretch}{1.1}

\begin{tabular}{
    p{0.30\linewidth}
    p{0.61\linewidth}
}
\toprule
\textbf{Function}
&
\textbf{Use in this work}
\\
\midrule

Artifact preservation
&
Stores raw outputs, scores, logs, traces, and execution metadata
\\

Turn diagnostics
&
Identifies the first failed turn and benchmark-specific error type
\\

Run comparison
&
Aligns cases and identifies changes in outputs or evaluator labels
\\

Selective retry
&
Regenerates failed cases and recomputes the merged benchmark score
\\

Human adjudication
&
Stores expert verdicts separately from official evaluator labels
\\

Version control
&
Pins benchmark code, patches, datasets, and execution settings
\\

\bottomrule
\end{tabular}
\end{table}

\section{Experiments}
\label{sec:experiments}

We evaluate whether current tool-calling benchmarks measure agent capability or evaluator behavior. Our analysis addresses three questions: \textbf{RQ1} whether official benchmark labels agree with expert trace-level judgments; \textbf{RQ2} which deterministic evaluator failures cause disagreement; and \textbf{RQ3} whether LLM-judge benchmarks produce reproducible scores under repeated evaluation.

\subsection{Experimental Setup}

\paragraph{Benchmarks and agents.}
We audit four benchmark families spanning deterministic and LLM-based evaluation: BFCL v4, $\tau^2$-Bench Retail, LiveMCPBench, and MCP-Atlas. BFCL v4 evaluates function calls and simulator state using AST and state-based checks. $\tau^2$-Bench combines database-state verification with action, communication, and natural-language assertions. LiveMCPBench and MCP-Atlas use LLM-based scoring over trajectories, final responses, or reference claims. We evaluate \texttt{moonshotai/kimi-k2.6} on $\tau^2$-Bench Retail and \texttt{minimax/minimax-m2.7} on BFCL v4, LiveMCPBench, and MCP-Atlas.

\paragraph{Audit protocol.}
For each task, we retain the complete user prompt, model trajectory, tool calls, tool outputs, final response, official benchmark verdict, evaluator diagnostics, and available environment states. Let $y_i\in\{0,1\}$ denote the official label for task $i$ and $h_i\in\{0,1\}$ the expert trace-level judgment. We define disagreement as
\begin{equation}
    m_i = \mathbbm{1}[y_i \neq h_i],
\end{equation}
and the benchmark-level misalignment rate as
\begin{equation}
    \mathrm{Err}(\mathcal{B})
    =
    \frac{1}{N_{\mathcal{B}}}
    \sum_{i=1}^{N_{\mathcal{B}}} m_i.
\end{equation}
A false negative occurs when $y_i=0$ and $h_i=1$; a false positive occurs when $y_i=1$ and $h_i=0$.

\paragraph{Expert adjudication.}
Three independent expert annotators reviewed the complete execution traces, including the user request, tool calls, tool outputs, environment-state changes, and final response. Disagreements were resolved through adjudication. The review required 89 annotator-hours in total, corresponding to approximately 10.8 minutes per task.

\paragraph{Execution and artifact collection.}
All runs were launched and inspected through Harness Lab with benchmark-specific diagnostic logging enabled. Harness Lab preserved raw vendor outputs, official score files, inference traces, evaluator logs, and available environment states. Benchmark revisions, local patches, datasets, endpoint configurations, and execution metadata were versioned for each run. The system was used to identify first-failure turns in BFCL, inspect decomposed $\tau^2$ rewards, compare repeated LiveMCPBench executions, and store human adjudications separately from official labels. Benchmark runners were deployed in Google Cloud \texttt{us-central1}; evaluated models were accessed through OpenAI-compatible HTTP endpoints. We therefore report benchmark-side compute rather than model-serving hardware.

\begin{table}[t]
\centering
\small
\caption{Audit scale and evaluator--human agreement. Agreement is the fraction of official labels matching expert trace-level judgments.}
\label{tab:audit_scale}
\vspace{3pt}
\resizebox{\textwidth}{!}{%
\begin{tabular}{l|l|l|rrrr}
\toprule
\rowcolor{tblHeader}
\textbf{Benchmark}
&
\textbf{Evaluator}
&
\textbf{Agent}
&
\textbf{Audited}
&
\textbf{Agreement}
&
\textbf{Misaligned}
&
\textbf{Error rate}
\\
\midrule

\rowcolor{tblGroupA}
$\tau^2$-Bench Retail
&
DB hash + NL assertions
&
Kimi-K2.6
&
112
&
\textbf{90.0\%}
&
11
&
9.8\%
\\

\rowcolor{tblGroupA}
BFCL v4
&
AST + simulator state
&
MiniMax-M2.7
&
200
&
80.0\%
&
40
&
20.0\%
\\

\midrule

\rowcolor{tblGroupB}
MCP-Atlas
&
LLM claim coverage
&
MiniMax-M2.7
&
89
&
87.0\%
&
12
&
13.5\%
\\

\rowcolor{tblGroupB}
LiveMCPBench
&
LLM judge
&
MiniMax-M2.7
&
95
&
\textbf{\textcolor{darkred}{69.0\%}}
&
\textbf{\textcolor{darkred}{29}}
&
\textbf{\textcolor{darkred}{30.5\%}}
\\

\midrule

\rowcolor{tblGroupC}
\textbf{\textcolor{darkgreen}{Total}}
&
--
&
--
&
\textbf{496}
&
\textbf{81.5\%}
&
\textbf{92}
&
\textbf{18.5\%}
\\
\bottomrule
\end{tabular}%
}
\end{table}

\subsection{RQ1: Do Official Labels Match Human Judgments?}

Table~\ref{tab:audit_scale} reports evaluator--human agreement across the four benchmark families. Among 496 audited tasks, 92 official labels disagree with expert judgments, yielding an aggregate misalignment rate of 18.5\%. The error rate ranges from 9.8\% on $\tau^2$-Bench Retail to 30.5\% on LiveMCPBench. BFCL v4 exhibits a 20.0\% error rate, while MCP-Atlas exhibits a 13.5\% error rate.

Misalignment occurs under both evaluation paradigms. Deterministic evaluators produce errors through brittle state comparisons, annotation defects, exact-match assumptions, and under-specified reward conditions. LLM-based evaluators produce errors through rubric instability, unsupported success judgments, answer-only scoring, and judge variance. These results indicate that replacing deterministic checks with LLM judging does not, by itself, produce reliable tool-calling evaluation.

\subsection{RQ2: Where Do Deterministic Evaluators Fail?}

\paragraph{BFCL v4.}
BFCL v4 exhibits 40 evaluator--human disagreements across 200 audited tasks. In the inspected 50-task multi-turn base export, the official evaluator assigns 25 passes and 25 failures. Of the 25 official failures, 20 are labeled \texttt{instance\_state\_mismatch} and 5 are labeled \texttt{empty\_turn\_model\_response}; thus, state mismatch accounts for 80\% of failures in this export.

Trace inspection shows that \texttt{instance\_state\_mismatch} conflates genuine task failures with evaluator artifacts. Genuine failures include omitted required actions. Artifact-driven failures include task-irrelevant state differences, punctuation-level discrepancies, full-object comparisons when only a subset of fields is relevant, and premature termination before a later corrective action. In such cases, the benchmark measures exact simulator-state conformity rather than whether the user objective was achieved.

\paragraph{$\tau^2$-Bench Retail.}
The $\tau^2$-Bench Retail audit contains 112 tasks and 11 disagreements. Eight are false negatives and three are false positives. False negatives arise when the agent completes the requested action but fails a brittle database-hash or communication assertion. False positives arise when unchanged database state or an incomplete rubric permits an unsuccessful trajectory to pass.

A representative false negative is Task~7. The database state and all five expected tool actions match, but the task fails because the final response does not contain the substring ``1628''. The user asks for the cost of remaining flights after two reservations are being canceled; the agent excludes the canceled reservations and returns \$708. The evaluator expects \$1,628, which includes flights no longer relevant to the request. The failure is therefore caused by an incorrect semantic target implemented as a substring check.

A representative false positive is Task~10. The user asks to return two orders, but the agent transfers the user to a human without invoking the return tool. Because the database remains unchanged and the rubric does not explicitly require the return action, the benchmark assigns a pass. This demonstrates that final-state scoring can reward inaction when the expected state is itself unchanged.

\begin{table*}[t]
\centering
\caption{Core audit statistics. BFCL reports the inspected 50-task multi-turn export, $\tau^2$ reports disagreement direction, and LiveMCPBench reports variability across repeated full runs.}
\label{tab:core_audit_statistics}
\vspace{2pt}
\scriptsize
\setlength{\tabcolsep}{3.5pt}
\renewcommand{\arraystretch}{1.08}

\begin{minipage}{0.31\textwidth}
\centering
\textbf{BFCL v4 failure mix}
\vspace{2pt}

\begin{tabular}{l|rr}
\toprule
\rowcolor{tblHeader}
\textbf{Outcome / Error} & \textbf{Count} & \textbf{Frac.} \\
\midrule
\rowcolor{tblGroupA}
Official pass & 25 & 50.0\% \\
\rowcolor{tblGroupA}
Official fail & 25 & 50.0\% \\
\midrule
\rowcolor{tblGroupB}
State mismatch & 20 & \textbf{80.0\%} \\
\rowcolor{tblGroupB}
Empty response & 5 & 20.0\% \\
\bottomrule
\end{tabular}
\end{minipage}
\hfill
\begin{minipage}{0.31\textwidth}
\centering
\textbf{$\tau^2$ disagreement direction}
\vspace{2pt}

\begin{tabular}{l|rr}
\toprule
\rowcolor{tblHeader}
\textbf{Error type} & \textbf{Count} & \textbf{Frac.} \\
\midrule
\rowcolor{tblGroupA}
False negative & 8 & \textbf{72.7\%} \\
\rowcolor{tblGroupA}
False positive & 3 & 27.3\% \\
\midrule
\rowcolor{tblGroupC}
\textbf{\textcolor{darkgreen}{Total}} & \textbf{11} & \textbf{100.0\%} \\
\bottomrule
\end{tabular}
\end{minipage}
\hfill
\begin{minipage}{0.31\textwidth}
\centering
\textbf{LiveMCPBench reruns}
\vspace{2pt}

\begin{tabular}{l|r}
\toprule
\rowcolor{tblHeader}
\textbf{Statistic} & \textbf{Value} \\
\midrule
\rowcolor{tblGroupA}
Tasks per run & 95 \\
\rowcolor{tblGroupA}
Valid reruns & 23 \\
\rowcolor{tblGroupB}
Minimum & 57.9\% \\
\rowcolor{tblGroupB}
Maximum & 76.8\% \\
\rowcolor{tblGroupB}
Mean & 69.4\% \\
\rowcolor{tblGroupB}
Std. dev. & 5.4 pp \\
\midrule
\rowcolor{tblGroupC}
\textbf{\textcolor{darkgreen}{Spread}} & \textbf{18.9 pp} \\
\bottomrule
\end{tabular}
\end{minipage}

\vspace{-4pt}
\end{table*}

\subsection{RQ3: How Reproducible Are LLM-Judge Benchmarks?}

We evaluate LiveMCPBench reproducibility using 23 full runs of the same 95-task configuration. Let $s_k$ denote the aggregate score from run $k$. We report the mean
\[
\mu = \frac{1}{K}\sum_{k=1}^{K}s_k,
\]
the standard deviation
\[
\sigma =
\sqrt{
\frac{1}{K-1}
\sum_{k=1}^{K}(s_k-\mu)^2
},
\]
and the score spread
\[
\Delta = \max_k s_k-\min_k s_k.
\]

Across 23 runs, LiveMCPBench scores range from 57.9\% to 76.8\%, with mean 69.4\%, standard deviation 5.4 percentage points, and spread 18.9 percentage points. The best and worst runs differ by 18 successful tasks out of 95. This variation is large relative to many reported leaderboard margins and makes a single-run score insufficient for fine-grained model comparisons.

The observed variance combines two sources. First, model trajectories vary across complete reruns. Second, the evaluator is itself stochastic: it regenerates task-specific key points and applies an LLM judge to the resulting rubric. Consequently, two evaluations may assess the same task under different generated criteria. The reported benchmark score is therefore a function of both agent sampling and evaluator sampling.

The audit also motivates a separation between benchmark design and benchmark infrastructure. Tool-Veritas applies deterministic checks to observable task outcomes and restricts LLM judging to qualitative criteria. Harness Lab preserves the execution evidence required to inspect those decisions, compare repeated runs, and record human corrections. Together, these components make evaluator behavior explicit rather than treating scoring as an opaque final stage.

\subsection{RQ4: Does Deterministic-First Evaluation Improve Human Agreement?} \label{sec:rq4}

We evaluate Tool-Veritas on 70 tasks with six models and compare its benchmark verdicts against expert trace-level judgments. Table~\ref{tab:custom_bench_results} reports deterministic-gate completion, final benchmark success, human success, and evaluator--human agreement. Agreement ranges from 91\% to 100\% across models, with an aggregate agreement of 95.5\% (401/420 model--task evaluations). This exceeds the agreement observed in our audits of BFCL v4 (80.0\%), $\tau^2$-Bench Retail (90.0\%), MCP-Atlas (87.0\%), and LiveMCPBench (69.0\%). All 19 Tool-Veritas disagreements are false negatives in which the benchmark is stricter than the expert judgment; we observe no false positives in which the benchmark passes an execution rejected by the human reviewer. In these evaluated settings, deterministic state gates with restricted LLM fallback substantially reduce unsupported passes while retaining high agreement with expert judgments. Because these results use different task distributions and model configurations from the audited benchmarks, we interpret them as evidence for the evaluator design rather than as a controlled benchmark-level ranking.

\begin{table*}[t] \centering \small \caption{Tool-Veritas results over 70 tasks per model. Agreement compares the final benchmark verdict with expert trace-level judgment. Strict disagreements are benchmark failures judged successful by the expert; lenient disagreements are benchmark successes judged unsuccessful.} \label{tab:custom_bench_results} \setlength{\tabcolsep}{4pt} \renewcommand{\arraystretch}{1.1} \begin{tabular}{lrrrrrrrr} \toprule \textbf{Model} & \textbf{Gates} & \textbf{Bench. pass} & \textbf{Pass (\%)} & \textbf{Human pass} & \textbf{Agree} & \textbf{Disagree} & \textbf{Lenient} & \textbf{Strict} \\ \midrule Sonnet-4.6 & 66 & 66 & 94 & 66 & 70 & 0 & 0 & 0 \\ GLM-5 & 63 & 62 & 88 & 64 & 68 & 2 & 0 & 2 \\ Kimi-K2.6 & 61 & 61 & 87 & 64 & 67 & 3 & 0 & 3 \\ MiniMax-M3 & 62 & 62 & 88 & 66 & 66 & 4 & 0 & 4 \\ GPT-4o-mini & 58 & 58 & 82 & 64 & 64 & 6 & 0 & 6 \\ Llama-3.1-8B & 38 & 37 & 52 & 41 & 66 & 4 & 0 & 4 \\ \midrule \textbf{Aggregate} & \textbf{348} & \textbf{346} & \textbf{82.4} & \textbf{365} & \textbf{401} & \textbf{19} & \textbf{0} & \textbf{19} \\ \bottomrule \end{tabular} \end{table*}

\section{Conclusion}

Tool-calling benchmarks are increasingly used as proxies for agent capability, but this paper shows that their evaluators are often not reliable enough to support that role. Across 496 expert-reviewed tasks from BFCL v4, $\tau^2$-Bench, LiveMCPBench, and MCP-Atlas, we find 92 evaluator-human disagreements, corresponding to an 18.5\% misalignment rate. These failures are not confined to a single benchmark or scoring paradigm. Deterministic evaluators fail through brittle state matching, trajectory lock-in, annotation errors, substring checks, and reward-basis mismatch; LLM-judge evaluators fail through rubric drift, hallucinated completion, answer-only scoring, and substantial run-to-run variance. In LiveMCPBench, repeated evaluation of the same setup changes the aggregate score by 18.9 percentage points, large enough to alter leaderboard conclusions.

In addition to identifying evaluator failures, we introduce two components for more auditable tool-calling evaluation. Tool-Veritas separates deterministic outcome verification from optional qualitative judging and records whether an agent succeeds initially or after a bounded repair. Harness Lab provides versioned benchmark execution, raw artifact preservation, case- and turn-level diagnostics, repeated-run comparison, selective retry, and human adjudication. These components implement the central recommendation of this work: benchmark evaluators should be inspectable, reproducible, and evaluated against human judgments of task success.

\section*{Availability}

We will release the trace-level audit artifacts, corrected benchmark components, Tool-Veritas configurations, and Harness Lab source code. Harness Lab is being prepared for release under the MIT License. Repository links and versioned artifact identifiers will be included in the final version.

 \bibliographystyle{plainnat}
\bibliography{main}

```latex
\appendix

\section{Qualitative Evaluator Failure Cases}
\label{app:qualitative_examples}

This appendix provides representative cases underlying the aggregate disagreement and reproducibility results in Section~\ref{sec:experiments}. The examples illustrate two distinct failure modes: an incorrect deterministic communication target in $\tau^2$-Bench and instability induced by stochastic rubric generation and judging in LiveMCPBench.

\subsection{$\tau^2$-Bench: Incorrect Communication Target}
\label{app:tau2_task7}

Table~\ref{tab:tau2_task7} summarizes a false negative from $\tau^2$-Bench Retail. The official evaluator assigns a failure, whereas expert review assigns a pass. The database hash matches the reference state and all five expected tool actions are completed correctly. The only failed component is a natural-language assertion requiring the final response to contain the substring ``1628.''

The user requested that reservations \texttt{XEHM4B} and \texttt{59XX6W} be canceled and asked for the total cost of any \emph{other} upcoming flights. The remaining reservations were \texttt{7WPL39}, costing \$402, and \texttt{3EMQJ6}, costing \$306. The agent therefore reported
\[
\$402 + \$306 = \$708.
\]
The reference target of \$1,628 includes reservations that the user explicitly asked to cancel. The evaluator thus rejects a task-consistent answer because its communication assertion encodes an incorrect semantic target.

\begin{table}[t]
\centering
\small
\caption{Representative $\tau^2$-Bench false negative.}
\label{tab:tau2_task7}
\setlength{\tabcolsep}{5pt}
\renewcommand{\arraystretch}{1.1}
\begin{tabular}{p{0.28\linewidth}p{0.63\linewidth}}
\toprule
\textbf{Field} & \textbf{Observation} \\
\midrule
Official verdict & Fail \\
Expert verdict & Pass \\
Database state & Matches the expected final state \\
Tool execution & All five expected actions match \\
Failed check & Final response does not contain ``1628'' \\
Agent answer & \$708 for the two remaining upcoming reservations \\
Reference target & \$1,628, including reservations selected for cancellation \\
Failure category & Incorrect ground truth implemented through a brittle substring assertion \\
\bottomrule
\end{tabular}
\end{table}

The agent first identified the two remaining upcoming reservations and reported their combined cost as \$708. After completing the requested upgrade and cancellations, it repeated the same total in the final response. The task therefore demonstrates that agreement on database state and tool execution does not guarantee a correct benchmark verdict when the communication target itself is semantically mis-specified.

\subsection{LiveMCPBench: Stochastic Rubric and Judge Instability}
\label{app:livemcp_instability}

LiveMCPBench exhibits a different failure mode. Repeated execution of the default evaluation pipeline on the same 95-task configuration produces aggregate scores ranging from 57.9\% to 76.8\% across 23 valid runs. The mean is 69.4\%, the standard deviation is 5.4 percentage points, and the total spread is 18.9 percentage points. This variation is large enough to change conclusions when leaderboard differences are only a few points.

\begin{table}[t]
\centering
\small
\caption{LiveMCPBench variability across 23 full runs of the default pipeline.}
\label{tab:livemcp_appendix_variance}
\setlength{\tabcolsep}{6pt}
\renewcommand{\arraystretch}{1.1}
\begin{tabular}{lr}
\toprule
\textbf{Statistic} & \textbf{Value} \\
\midrule
Tasks per run & 95 \\
Valid runs & 23 \\
Minimum score & 57.9\% \;(55/95) \\
Maximum score & 76.8\% \;(73/95) \\
Mean score & 69.4\% \\
Standard deviation & 5.4 pp \\
Maximum--minimum spread & 18.9 pp \\
\bottomrule
\end{tabular}
\end{table}

The default OpenBench scorer performs two stochastic LLM calls for each task. First, \texttt{identify\_key\_points(task)} regenerates evaluation criteria from the task text. Second, \texttt{grader\_model.generate(...)} evaluates the trajectory and final response against the generated criteria and returns a success or failure verdict. Human-authored step annotations are used only as a fallback when key-point generation fails; under normal execution, they do not determine the rubric.

This design introduces two sources of evaluator variability. First, regenerated key points can differ in specificity and strictness across runs, causing the same underlying behavior to be assessed against different criteria. Second, the final LLM verdict is itself stochastic and can flip on borderline traces. The default judge is \texttt{gpt-4.1-mini}, rather than the stronger judge configuration recommended in the benchmark paper, which may further increase instability.

Because these are complete benchmark reruns, the observed spread combines variation in agent trajectories with variation in rubric generation and judging. It therefore measures end-to-end pipeline instability rather than isolated judge noise. A controlled judge-only analysis would require repeatedly rescoring fixed trajectories under fixed and regenerated rubrics.
```

\newpage

\end{document}